\documentstyle[multicol,aps]{revtex}
\begin{document}
\title{Elasticity of Entangled Polymer Loops: Olympic Gels}
\author{Thomas A.Vilgis and  Matthias Otto}
\address{Max-Planck-Institut f\"ur Polymerforschung, Postfach 3148,\\
D-55021 Mainz, Germany}
\date{\today }
\maketitle

\begin{abstract}

In this note we present a scaling theory for the elasticity of
olympic gels, i.e., gels where the elasticity is a consequence
of topology only. It is shown that two deformation regimes exist. The
first is the non affine
deformation regime where the free energy scales linear with the deformation. 
In the large (affine) deformation regime the free energy is
shown to scale as $F \propto \lambda^{5/2}$ where $\lambda$ is the
deformation ratio. Thus a highly non Hookian stress -  strain
relation is predicted.
\end{abstract}

\begin{multicols}{2}

In this communication we compute the scaling of the elastic modulus of an
olympic gel. This name has been created by de Gennes \cite{gennes}
since the configuration of such gels resemble very much the olympic rings.
Olympic gels are very peculiar elastic materials as their elasticity
does not come from crosslinks as in conventional rubbers. Actually ideal
olympic gels do not contain any crosslinks, but consist of concatenated
rings only. In this sense the elasticity of such materials
is expected to be very different from classical rubbers and strong deviations
from the non-Hookian deformation behavior \cite{neohook,edvil}
(as observed in the low deformation regime in conventional rubbers)
must be expected. Actually, the elasticity of such olympic gels has
not been calculated yet and we attempt to present a simple argument
in this issue. Apart from the case of synthetic materials these
considerations are important for biological systems too. It is
well known that highly entangled DNA rings exists and play an important
role in biology \cite{adams,wasserman,seeman,smith}

Unless like in conventional soft materials, such as rubbers
the precise calculation of the modulus of topological gels
is difficult because the exact topological state of the
gel must be known. Specifying topological states is a general
problem in polymer physics and has been first discussed in \cite{edwards67}.
The corresponding mathematical
problem is the classification of knots and links \cite{kauffman}.
Despite its incompleteness already noted in \cite{edwards67},
in general the Gauss invariant is used in polymer physics because
it is the most simple invariant that explicitly contains the
polymer conformation in contrast to algebraic invariants in
knot theory.
Even using the Gauss invariant,
it is, in general, not very simple to describe
the linking status of the network. Only for the
easy case of a non-concatenated
melt of rings this seems to be possible, but already here
many complications appear \cite{brervil}.

The synthesis and preparation
of such gels is also a very difficult task. Some of the
problems are already discussed in de Gennes' book \cite{gennes}.
The crucial point is that the ring closure of the chains
must be carried out at concentrations larger than the overlap
concentration $c^*$. The reaction process is carried out
in two steps: First a certain amount of rings is cyclicized
by end group reaction. Then further linear chains are added
to the given sample. These additional rings are then cyclicized
again. With variation of the concentration different
entanglement numbers can be expected. The solvent must then
be evaporated to get the bulk network. In such an ideal synthesis
the topological state of the network depends on the concentration
of preparation.  One limit is the cyclizaton in the
linear melt state (a {\em gedanken} experiment).
Such prepared topological gels will be important for motivation
of the following scaling analysis. Assume therefore a condensed melt
of linear chains. It is well known, that the excluded volume forces
are screened out \cite{gennes,doi}. The screening of the excluded volume
can also be asserted to the strong interpenetration of linear
(= one dimensional connectivity) chains \cite{haronska}.
As a result, the chains
behave gaussian in the melt. If then each long chain is closed to a ring
({\em in Gedanken}), the size of the ring polymer would not change.
The reason for this is, that each ring is (in average) concatenated
by $n \simeq \varrho \ell^3 N^{1/2}$, where $\varrho$ is the
density of the melt, $\ell$ the typical size of a monomer, and
$N$ the degree of polymerization. In the limit of
scaling, $n$ can be identified by the winding number.
Such natural conjectures have been
also mentioned by Cates et al. \cite{cates}. For a more rigorous
definition of the winding number in such gel-like systems
we refer the reader to our previous publication \cite{brervil}.
In this paper we had already confirmed the scaling conjecture presented
first by Cates et al. \cite{cates} by a careful treatment of the
Gaussian linking number. The main result there was to show that the
conformation of a ring in an
unlinked melt of rings is naturally non - gaussian, i.e., the
size $R \propto N^{\nu}$, where $\nu < 1/2$.

In the present note we treat dense {\em linked} olympic gels without solvent. 
The discussion
of solvent free gels yields the "bare" scaling of the elastic modulus
without additional contributions from the excluded volume and
swelling.

To do this, we have to use non-conventional arguments, because
most of the theories for ideal, ordinary rubber networks are
"single chain theories", i.e. all the elasticity of the network
is computed by the contribution of the
elasticity of one single network strand, which
is then multiplied by the number of
elastically active chains \cite{edvil}.
These well known results can be summarized by the free energy of the
network as function of deformation $F \propto Mk_{\rm B}T \lambda^2$,
where $M$ is the number of active chains and $\lambda$ is the deformation
ratio of the chain.
Although this seems to be
very simplified, the results are in reasonable agreement with
low deformation experiments and recent simulations \cite{kremer}, as long as 
higher order effects at large deformations
are ignored. In this note we put
forward a scaling theory in a similar spirit. To do this we estimate the
single ring behavior and conclude from this basis the elasticity of
the entire gel. For the subsequent analysis we need the following
assumptions
\begin{itemize}
\item{The rings are not self-knotted}
\item{The topological state of the network can be described by a
global winding number $n$, which on the average is the same for each ring
in the network}
\item{The direct influence of excluded volume effects are ignored}
\end{itemize}
Again, these assumptions seem to be crude, but these turn out to be
sufficient for the type of analysis presented here. A mathematical
formulation of the problem of elasticity of the olympic gel
will be presented elsewhere \cite{tobe}.

In the following we start from a Flory type estimate of the
free energy of an entangled ring in a olympic gel, where the
average winding number of each ring is assumed to be $n$.
The free energy of an entangled ring in a network can be
written as
\begin{equation}
\label{1}
F =k_{\rm B} T \frac{N}{R^2} + \mbox{const} \frac{R^3}{nN}
\enspace .
\end{equation}
Here the first term is the gaussian elastic part of the ring in the gel
and the second term represents the pressure experienced
by the ring that comes from all the surrounding rings.
If the ring is not entangled with the others the second term
would read $R^3/N$  obtained by replacing $n$ by $n+1$ in Eq.(\ref{1})
and $n=0$.
Note that in this case
the present {\em ansatz} for the free energy agrees
with the one proposed by Cates
and Deutsch \cite{cates} for non concatenated ring melts.
If, however, other polymers are entangled with
the ring under consideration they exert a repulsive force per winding number 
from inside and outside the ring, and
consequently they do not contribute to
the packing pressure. By this, we mean that the dense melt of rings is as
closely packed as possible. Thus one ring experiences a pressure induced by
the surrounding ones. The packing term is, however,  reduced by the average
winding number. The factor $1/n$ in front of the second term
has its reason in the "screening" of the packing pressure. Each other
ring, that is entangled with the ring polymer under consideration
reduces the packing pressure.
In the limit of the cyclization in the melt the free energy
contribution must be of the order of ${\cal O}(1)$, since
the rings must be gaussian. Note that the free energy described
by Eq.(\ref{1}) does not contain an upper critical dimension.
The interaction term is therefore important in all spatial
dimensions. Instead of an upper critical dimension
it contains a limiting winding number such that $R \propto \sqrt{N}$,
corresponding to the melt cyclization {\em gedanken}experiment.
Minimization of the
free energy with respect to $R$ yields the size for the
ring in the network
\begin{equation}
\label{2}
R_N \propto n^{1/5}N^{2/5}
\enspace ,
\end{equation}
where all (to the purpose of this paper) irrelevant constants have been
dropped. The latter equation has interesting consequences. For small $n$
the scaling results agree with those proposed by Cates et al. \cite{cates}.
The rings appear compressed in the melt. Similar findings have been put forward
by a  more rigorous theory by us recently \cite{brervil} and by numerical
simulations of the problem \cite{weyer,wittmer}. On the other hand, if
the mean winding number is close to that given by the average density of
the system, i.e., $n \propto \sqrt{N}$, the ring is gaussian at
melt  cyclization conditions. In this case the topological effects are screened
and the ring finds itself in a natural melt environment. If the winding number
is larger, for example $n \propto N$ than too many rings are connected with
each other and the ring configuration stretches out. So far the requirements
have been satisfied by the simple scaling {\em ansatz}.

The next task is now to find the asymptotic distribution function for
the typical size $r$ of the ring in the gel.
A reasonable form of the distribution function
is given by $P(r,N,n) = {\cal N} r^{\theta} f \left( \frac{r}{R_N}\right)$,
where ${\cal N}$ is a suitable but uninteresting normalization.
$\theta$ determines the short distance behavior and the function
$f(x)$ has to be determined. In fact, for the elasticity the knowledge of
the scaling function $f$ is sufficient, because the power in front of the
the distribution function yields only irrelevant (logarithmic)
corrections. To find the appropriate
distribution function, we start from the
asymptotic behavior of the scattering function
defined by the size of the ring.
Then, by standard methods, such as steepest descent Fourier
inversion
\cite{gennes} it is easily found that the asymptotic form
of the distribution is given by
\begin{equation}
\label{3}
P(r,N,n) \propto \exp \left\{ - \left( \frac{r}{R_N} \right)^{5/3}\right\}
\enspace ,
\end{equation}
This asymptotic form will be sufficient in the scaling limit. This is indeed
the key equation of the paper. Together with the equation \ref{2} it yields
the correct asymptotic behavior of the distribution function. For low values
of the average winding number $n$ it contains the limits suggested in
\cite{cates}

The olympic gel is considered to be entropy elastic. Thus we may
conclude that the elastic free energy of the typical ring in the
olympic gel as a function of its elongation $r$ is given by
\begin{equation}
\label{free}
F_{\rm s} (r) \propto T {1 \over (nN^2)^{1/3}} r^{5/3}
\enspace .
\end{equation}
The tension is given by the derivative and thus we find a non linear
force extension relationship due to the non gaussian structure of the
rings in the network, i.e.,
\begin{equation}
f = T  {1 \over (nN^2)^{1/3}} r^{2/3}
\enspace .
\end{equation}
The latter result is the force extension law for one ring in the
olympic gel, and corresponds to the according single chain deformation law
in conventional rubber theory, i.e., $f = (T/N)r$ \cite{neohook}.
Therefore (non-)Hookian deformation behavior in olympic gels
cannot be expected.

At this point we have the possibility to observe two different
deformation processes and regimes.
At low deformations
the entanglements do not act as severe constraints, but have
many degrees of freedom, similar to entanglements in conventional
rubbers. It has been shown that there the entanglements yield
a "softening" of the modulus, if compared to the classical gaussian
theory \cite{edvil,vilgis:86,ball:81}. This softening of the
modulus corresponds to the slippage and sliding of entanglements.
This happens if the deformation of the individual ring
in the gel is such that only the mean contour between two
entanglements, i.e., $N/n$ take part on the
deformation. This defines the deformation ratio
$\lambda_0 = (N/n)/(R_N) \cong (N/n^2)^{3/5}$. Note that $\lambda_0$
in the melt preparation conditions, when $n \propto \sqrt{N}$,
is ${\cal O}(1)$, which is physically sensible: At high degree
of entangling the topological constraints act immediately as
crosslinks from the lowest deformation. From the
arguments presented in ref. \cite{edvil,vilgis:86} the
maximum deformation can be estimated to $\lambda_{\rm max}
= {\cal O}(\sqrt{N/n})$. The latter is always less than $\lambda_0$ as long
as $n \leq N^{1/7}$. For the validity of the scaling arguments
this must always be the case here.

Consider first the low deformation regime
$\lambda < \lambda_{0}$.
In the low deformation regime of olympic gels the slippage of the
topological constraints dominates. The main problem with the low deformation
regime is that the relevant chain length is not fixed
\cite{edvil,vilgis:86,higgs}. To see this point consider conventional rubbers,
where the fixed length scale is given by the mesh size. In such olympic gels a
clear length scale (at least in the low deformation regime) cannot be defined,
because the system is ruled by a large number of degrees of freedom.

It has been shown that the
deformation process can then be described by an effective distribution
$\tilde P(r) = \int {\rm d}N p(N) P(r,N,n)$
function on the level of a single ring. As a consistent model we choose
for $p(N)$ the entanglements slack \cite{needs}, $p(N) \propto
\exp(-N/N_0)$, which has been successfully
applied to entanglement problems (see \cite{edvil} and references therein).
$N_0$ is a mean excursion of the ring form the most probable conformation.
The effective distribution is then given by the asymptotic form
\begin{equation}
\tilde P(r,N_0,n) \propto \exp \left(-\frac{r}{n^{1/5}N_0^{2/5}}  \right)
\enspace ,
\end{equation}
which is consistent with eq.(\ref{2}) since the mean size of the
ring is not altered. The macroscopic free energy of an
ensemble of $M_{\rm R}$ rings (per unit volume)
is then given by multiplying the above equation by the
number of constraints present. These are the number
of entanglements
$M_{\rm R}n$. To introduce the
deformation we replace $r$ by $\lambda R_N$ and average thus over the
conformation as in the simplest theories in classical networks.
The total free energy is
then estimated by
\begin{equation}
F_{} = T M_{\rm R}n \lambda^{}
\enspace ,
\end{equation}
The non-Hookian linear increase of the free energy is entirely due
to the large degrees of freedom of the constraints, and is
in some way similar
to the low deformation regime in highly entangled rubbers, with
strong entanglement sliding, when the sliplink
contribution becomes very weak (see ref. \cite{edvil} for details).
Consequently, the free energy of the olympic gel is
linear in the elongation $\lambda$, and the interesting
result is that the force needed for elongating the gel is constant, i.e.,
$f = T n M_{\rm R}$ in the low deformation regime.
In terms of macroscopic variables it is given by  $f=(T \varrho)/N$, where
$\varrho = M_{\rm R}N$ is the macroscopic density of the gel.
Such a deformation regime is not observable in
classical (highly entangled) gels,
because in olympic gels the degrees of freedom
of the non crosslinked polymers are much larger, leading to a very weak
solid. This is, of course, because olympic gels consist only of
entanglements.

At larger deformations ($\lambda > \lambda_0$)
the individual chains are deformed also,
and the topological constraints act as crosslinks. Therefore
the free energy must be proportional to the effective
umber of constraints, i.e., $M_{\rm R}n$.
In this
case we obtain
\begin{equation}
F \cong M_{\rm R} n \lambda^{5/3} = \varrho \frac{n}{N} \lambda^{5/3}
\enspace ,
\end{equation}
Obviously $M_{\rm R} n$ is the effective number of
crosslinks in the (affine) deformation regime, where the
topological constraints act almost as crosslinks.
The measured force $f$ is
then give by the derivative of the
free energy with respect to the deformation $\lambda$., i.e.,
$f \cong \varrho (n/N) \lambda^{2/5}$ and is larger by a factor of $n$
due to the number of constraints but much weaker in the deformation
dependence when compared to classical rubbers. In the latter case
the force is roughly given by $f = \varrho (1/N) \lambda$ \cite{neohook}.
The factor of $1/(N/n)$ in the force
can be indeed interpreted by the effective meshsize of the olympic gel.
distance

In summary, we have presented a simple calculation of the elasticity
of olympic gels as a classical example of weak solids. We found
two relevant deformation regimes which are determined by the
topological state of the network. The first is the non-affine regime
where the modulus is very weak indeed and the scaling is determined by
the average winding number and the degree of polymerization of the
rings. In the second deformation regime  the topological constraints act
similar
as crosslinks. Therefore the (low deformation $\lambda \approx 1$) modulus
is given by $G_{\rm affine} \cong \varrho (n/N)$.
When the mean winding number $n$ is of the order
of $\sqrt N$ then the modulus becomes larger compared to the classical
rubber. This case corresponds to ring closure of the polymers in the
melt state. This high modulus in the affine regime is naturally determined
by the large number of effective crosslinks. We expect that the present
results have some applications in biological systems, too.
\end{multicols}

\end{document}